\documentstyle[aps,epsfig,multicol]{revtex}

\newcommand{\figwidth}{0.48\textwidth}

\begin{document}
\draft

\title{Self-organized criticality in the Kardar-Parisi-Zhang-equation}
\author{G\'abor Szab\'o$^{1,2}$, Mikko Alava$^{1}$ \and J\'anos
Kert\'esz$^{2}$}
\address{$^1$Helsinki University of Technology, Laboratory of Physics,
02105 HUT, Finland}
\address{$^2$Budapest University of Technology,
Department of Theoretical Physics, H-1111 Hungary }
\date{\today}

\maketitle

\begin{abstract}
\noindent Kardar-Parisi-Zhang interface depinning with quenched noise
is studied in an ensemble that leads to self-organized criticality in
the quenched Edwards-Wilkinson (QEW) universality class and related
sandpile models.  An interface is pinned at the boundaries, and a
slowly increasing external drive is added to compensate for the
pinning. The ensuing interface behavior describes the integrated
toppling activity history of a QKPZ cellular automaton. The avalanche
picture consists of several phases depending on the relative
importance of the terms in the interface equation.  The SOC state is
more complicated than in the QEW case and it is not related to the
properties of the bulk depinning transition.
\end{abstract}

\begin{multicols}{2}[]

One of the more interesting developments in non-equilibrium
statistical physics has been the realization that inhomogeneous
systems give rise to new, rich behavior compared to those in which
translational invariance is present. A prime example thereof is the
criticality found in so-called sandpile models, coined
'self-organized' due to the absence of an obvious control
parameter~\cite{btw}.

Usual 'self-organized critical' SOC sandpiles have two central
features. First, the system is driven very slowly so that individual
avalanches occur on a separate, fast time-scale. Second, the drive
(like the addition of grains) and the dissipation (loss of grains)
occur non-uniformly. Typically the system is forced, on the average,
in an uniform fashion but open boundaries create dissipation, in
sandpiles via the loss of grains. These two characteristics are the
keys to describe, and the earmarks of SOC (for early work see
\cite{various}) \cite{new-av,oma}.

Essential to a wide class of SOC models is that the movement of grains
has a Laplacian character. Consider now the {\em history} of activity,
topplings in such models. The activation of a site means that it has
received enough grains, $z(x,t)$, from outside or through
redistribution, to overcome a local activation threshold: $z(x,t) >
z_c(x,t)$. For a large class of models, this implies that if $H(x,t)$
defines the time-integrated activity, $H = \int^t \rho(x,\tau) d\tau =
\int^t \Theta(z(x,\tau)-z_c(x,\tau)) d\tau$, then after discretization
$H$ obeys the discrete {\em interface equation}
\begin{eqnarray}
        H(x,t+1)&  = &\left\{\begin{array}{ll}
                        H(x,t) + 1,  &  f(x,t) >0,  \\
                        H(x,t),      &  f(x,t) \le 0,
                   \end{array} \right. 
\end{eqnarray}
where $f$ denotes the local 'force' acting on the interface
\cite{oma}. This can be recast in the form
\begin{equation}
	\frac{\Delta H}{\Delta t}=\Theta \left(
	        \nu \nabla^2 H + \eta(x,H) + F(x,t) -F_c \right) .
                           \label{eq:H-def}
\end{equation}
The Laplacian comes from the average redistribution of grains, and a
constant $\nu$ takes into account the specific updating rules of the
system.  $\eta$ is a {\em quenched noise term} that originates from
the random rules of sandpiles. $F(x,t)$ measures the external drive
and changes on the slow-time scale, and $F_c$ is the toppling
threshold written as a force acting on the interface. By casting the
dynamics of a sandpile in the form of a driven interface in a random
enviroment ($\eta$) one sees that the SOC critical point is reached
under the combined action of a slowly-increasing external force
$F(x,t)$ and the boundary conditions, $H=0$, which amount to the loss
of grains. This is a slightly unusual case of a {\em depinning
transition} of the quenched Edwards-Wilkinson universality class,
since the macroscopic shape of the interface is parabolic (in 1D)
\cite{oma}. The avalanche picture in an ordinary depinning transition
\cite{nattermann-etal:1992,leschhorn:1993,narayan-fisher:1993,chauve}
is however different since the translational invariance is here broken
as manifested already in the average interface profile - which tells
us that the activity density peaks in the bulk, in the center of the
system.

The connection of SOC to non-equilibrium phase transitions can be made
both in the interface context and in the realm of absorbing state
phase transitions, exemplified by the contact process and various
reaction-diffusion models \cite{new-av,rd}.  One way to classify SOC
models is to consider the noise correlator in the presentation at
hand.  Such analysis however assumes certain symmetries included in
the representation (the QEW Langevin equation; the Langevin equation
describing the active site density). A pertinent question thus arises:
what happens if the ensemble, in which SOC is observed, is varied with
respect to its fundamental symmetries?  In this work we illustrate
this by analyzing the outcome for the Kardar-Parisi-Zhang universality
class. The difference between the QKPZ and the EW equations (or a
non-linear Langevin equation compared to a linear diffusion equation)
is that the diffusion equation is insensitive to most initial profiles
due to the fundamental symmetries of the equation, while in the QKPZ
case the growth component perpendicular to the interface slope is
always non-zero. This means that there will be a new history
dependence in the QKPZ interface which is absent from the EW dynamics
and from usual absorbing state phase transitions.

Consider, in analogy to the sandpile interface representation, the
quenched KPZ equation \cite{kpz} which reads
\begin{eqnarray}
\frac{\Delta H}{\Delta t}
= \Theta( \nu \nabla^2 H +
{\lambda \over 2} (\nabla H)^2 +  \eta({\mathbf x}, H) + F(x,t)),
\label{QKPZ}
\end{eqnarray}
where the nonlinearity proportional to $\lambda$ has been added to the
QEW equation. Note the step function that enforces the constraint that
the interface cannot propagate backwards. In the case of the sandpile
equation, it also discretizes the interface velocity to zero or unity
which in the absence of mixing by randomness creates an important
Abelian symmetry and thus becomes crucial in that case \cite{dhar}.
The external drive is in the following taken to be uniform in space,
$F(x,t) \rightarrow F(t)$ and the noise to be quenched and
delta-correlated, $\langle \eta({\mathbf x}, H) \eta({\mathbf x'}, H')
\rangle = 2D \delta^d ({\mathbf x} - {\mathbf x'}) \delta(H-H')$.
Equation (\ref{QKPZ}) has to be augmented with suitable boundary
conditions in order to reach a steady-state. As in the QEW or sandpile
ensemble, we use the selection $H(x,t)=0$ at $x=0$, $x=L+1$ (in 1+1
dimensions) instead of e.g.  periodic boundary conditions. In the case
of such ordinary boundary conditions Eq.~(\ref{QKPZ}) has a depinning
transition at a critical $F_c$. The exponents depend on whether
$\lambda$ approaches zero with $\Delta F \equiv F - F_c \rightarrow 0$
or not, and in the latter case we have a true QPKZ depinning
transition \cite{kpz,toinen}.

An interface evolving from a flat boundary condition will develop a
non-zero roughness and finally stops due to the influence of boundary
conditions.  The roughening, the development of average local slopes,
depends on the sign of $\lambda$.  To reach SOC-like conditions, the
separation of timescales, or a vanishingly small but positive
interface velocity (average activity level) $v = 0^+$ one has to
choose the right way of driving the system, $F(t)$. For $F(t)>0$ the
interface becomes {\em unstable} if $\lambda>0$ and ${dv \over dt}
>0$. This breaking of up/down symmetry, absent in the continuum QEW
model, shows the first sign that the possible SOC state in the QKPZ
equation is different.  If $\lambda <0$ it becomes posssible to take
the SOC limit of vanishingly slow external drive,
\begin{equation}
F(t) = \zeta t, \,\,\,\,\,\, \zeta \rightarrow 0
\end{equation}
from which follows that a separation of time-scales happens, similarly
to the QEW or normal sandpile case. In the following we investigate
the ensuing interface dynamics.

To this end, a cellular automaton \cite{lekpz} version of
Eq.~(\ref{QKPZ}) is defined as follows. Let $f_i$ be a microscopic
force at the discretized sites $x \rightarrow i$, such that for
$f_i>0$ $i$ topples. $f_i$ reads in 1D
\begin{eqnarray}
f_i(t) = \nu \left[ H_{i+1}(t)+H_{i-1}(t)-2H_i(t) \right]
\nonumber \\
+ {\lambda \over 4}
\left( [H_{i}(t)-H_{i-1}(t)]^2 + [H_{i+1}(t)-H_{i}(t)]^2 \right)
\nonumber \\
 + F(t) + \eta_{i, H_i}.
\label{KPZdiscr}
\end{eqnarray}
This can be interpreted, in analogy to SOC ones, as a {\em QKPZ
sandpile} where a site gives $2\nu$ particles or units of energy to
its neighbors once it topples, and the toppling criterion also
compares the integrated activity at site $i$ to its neighbors. In the
interface picture $H_i(t+1)=H_i(t)+\Theta(f_i(t))$. With this
definition, $H(x,t) = \int^t \rho(x,\tau) d\tau$ maps the discrete
QKPZ equation to the 'sandpile' and vice versa. Notice that the
nonlinearity, proportional to $\lambda$, above is discretized in a
particular way, to avoid the numerical instabilities of the QKPZ
equation in a reasonable fashion \cite{numer}.

\begin{figure} \begin{center}
\includegraphics[width=\figwidth]{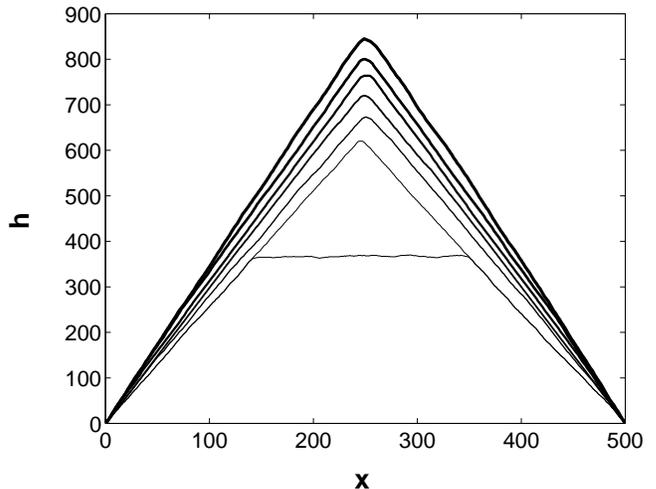}
\caption{Typical interface configurations for a 1D system of 500
sites. The one with the flat top shows an unstable one from the
initial growth event before the first 'triangle' is created. The other
configurations are stable, from every 20000th time step.}
\label{snapshot}
\end{center} \end{figure}

Fig.~\ref{snapshot} shows a series of snapshots from a simulation of
the QKPZ equation with $\eta_{i, H_i} =\pm g$ with equal
probability. This describes random toppling thresholds that change
after each toppling in the sandpile language. The system is driven in
such a way that $F(t)$ is increased after an avalanche has relaxed
(defined in the usual way by the stoppage of all activity), so that
one site becomes active, and the interface advances. It is evident
that the dynamics of such a system consists of at least of the
development of an initial profile, and the subsequent phase(s). The
macroscopic shape of the QKPZ interface is to first order triangular
in 1+1-D as expected with $\lambda<0$.  The slope follows from the
different terms in Eq.~(\ref{QKPZ}).  For a flat slope the surface
tension is negligible, and the slope arises from the balance of the
nonlinearity and the driving force. This gives $\langle |\nabla h|
\rangle = \sqrt{2 \zeta t/\lambda}$. The development of such a
profile, determined by the symmetries of the interface equation, is
analogous to the development of the steady-state grain configuration
in ordinary SOC sandpiles from an empty system.  Next we discuss the
phase that appears after the initial transient.

\begin{figure} \begin{center}
\includegraphics[width=\figwidth]{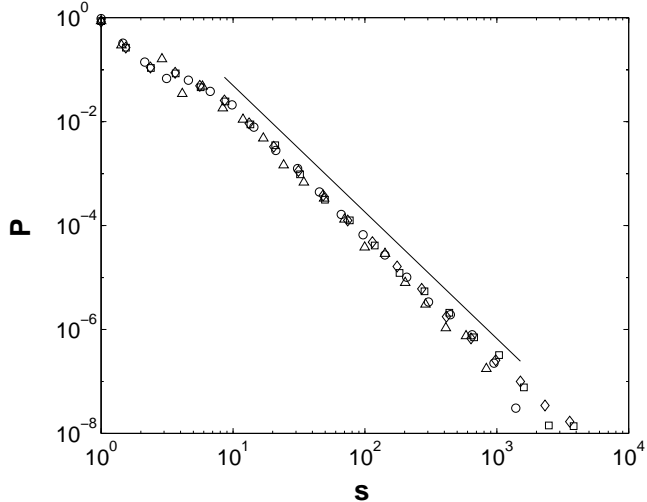}
\caption{Probability distribution function of the avalanche sizes for
1D ensembles. System sizes are 250, 500, 750, and 1000 for
{$\triangle$}, {$\circ$}, {$\Box$}, and {$\Diamond$}, respectively.
The parameters are $\nu=20, \lambda=-4, g=10$. Dynamics was only
followed for configurations with average interface slopes less than 10
(see text). The parallel line indicates the region of power law
fitting resulting in an exponent of $-2.5 \pm 0.1$.}
\label{avalsize_1d_autom}
\end{center} \end{figure}

As in ordinary SOC, avalanches are characterized by the duration $T$,
area $l$, and number of topplings $s$ (the number of times sites
advance). The initial growth, giving rise to the triangular shape, is
neglected. We obtain for the avalanche sizes in this regime the result
shown in Fig.~\ref{avalsize_1d_autom} for four system sizes.  The size
distribution has \emph{asymptotic power law decay} with a measured
exponent of $\tau_s = 2.5 \pm 0.1$ with $P(s) \sim s^{-\tau_s}$.  Due
to the large $\tau_s$ exponent the average avalanche size is not
dependent on $L$, unlike in usual SOC sandpile models.

\begin{figure} \begin{center}
\includegraphics[width=\figwidth]{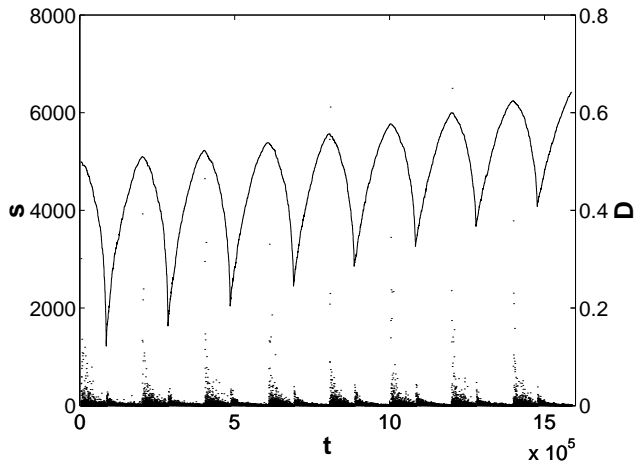}
\caption{Avalanche sizes $s$ versus time and the standard deviation of
the local slope $D$. $\nu=20$, $\lambda=-4$, $g=10$ and the system
size is 1000. The correlation between the two quantities is obvious.}
\label{avalevul}
\end{center} \end{figure}

However, the avalanches are time-translation invariant only {\em in
the average sense}, due to the role of the average interface slope
$m$.  The avalanche sizes are plotted versus time on
Fig.~\ref{avalevul}. On the same diagram, the standard deviation $D$
of the local slope is also shown.
The largest avalanches take place when the roughness of the interface
$D$ is the largest. Smaller, but still relatively large avalanches
develop where the sides of the triangular interface are smooth. The
local slopes, i.~e. $|H_i-H_{i-1}|$ alternate between $n$ and $n+1$,
where $n$ is a positive integer. It is not surprising that a smooth
interface evokes large avalanches, because there are no fluctuations
along the surface that would stop an upwards propagation of the
avalanche.  When $D$ is the smallest, the first few avalanches start
from either bottom of the triangle and propagate up to the top,
increasing the interface height by one at every site.  The really
large avalanches occur at the maxima of $D$.  They typically do not
increase the average gradient of the side, but only the height of
interface evenly along the side. In spite of the extra
slope-dependence, there is no characteristic size as is typical of SOC
and the correlation length equals the system size. The period of the
avalanches is simply related to the squared system size. This is since
the area covered by all the avalanches between two integer values of
$m$ is proportional to $L^2$, and, the average avalanche size and
duration are of the order of unity: $\langle s \rangle \sim {\cal O}
(1)$ and $\langle T \rangle \sim {\cal O} (1)$.

The support or the area of the avalanche is as usual defined by the
number of sites that have toppled at least once during an
avalanche. Notice that an avalanche always tends to spread upwards,
and thus in 1D the starting and final, extremal sites usually give the
area directly. Sometimes, the Laplacian makes it so that an avalanche
can propagate over the top of the 'triangle'.  We find that the
probability of starting an avalanche linearly increases with the
location of a site and the probability of an avalanche to stop grows
linearly also as the top is approached.

The area of the avalanches, defined as $l=i_r-i_l+1$, is expected to
follow a similar power law as the avalanche sizes do and one obtains
that $\tau_l \sim 3.0 \pm 0.1$.  The avalanche lifetime $T$ is the
time that elapses between two successive stationary states in the
system and also behaves as a power-law, $P(T) \sim T^{-2.9 \pm
0.2}$. Due to the fact that the average avalanche properties are
defined by the small avalanches no analogy of the usual scaling
relations between the exponents seems to exist.

The above results have, however, been obtained in just a window of
time. Consider an interface (side of the triangle) with an uniform
slope $m$ and an avalanche which has just started to propagate and
reached site $i$. The local force at site $i$, at time $t$, becomes
$f_i(t)=\nu(m-1-m)+{\lambda \over 4}\big[(m-1)^2+m^2\big]+F+\eta_i2
 = -\nu+{|\lambda| \over 4}(2m-1)-{|\lambda| \over 4}2m^2+F+\eta_i$
and at $i+1$ $f_{i+1}(t)=\nu[m-(m-1)]+{\lambda \over
4}\big[m^2+(m-1)^2\big]+F+ \eta_{i+1}
 =  \nu+{|\lambda| \over 4}(2m-1)-{|\lambda| \over 4}2m^2+F+\eta_{i+1}.$
Assuming that on average on an even interface $F$ balances the local
forces for slope $m$ ($F \approx {|\lambda| \over 4}2m^2$) and
neglecting the noise fluctuations, we find a critical slope
\begin{equation}
m_{cr} \approx {{2\nu} \over {|\lambda|}}+{1 \over 2}.
\end{equation}

This means that the cell at $i$ may stay active for the next updating
step at time $t+1$ when $m > m_{cr}$.  The natural progress of
avalanches is now such that the site at $i+1$ is normally always
activated if the interface is smooth enough. In this phase the site at
$i+1$ gets active and topples at $t+1$, but at the same time $H_i
\rightarrow H_i+1$ as well. Thus, a short region of the interface with
a smaller gradient will be formed and starts to propagate upwards,
giving rise to very large avalanches.  Since these are more difficult
to induce, the majority of the avalanches are of size unity, and tend
to smoothen out the interface fluctuations. The avalanche statistics
data presented previously is restricted to systems with a maximum
slope of 10: the critical slope for those systems is $m_{cr} \approx 2
\cdot 20 / 4 + 1/2=10.5$. In this regime above the critical slope
$m_{cr}$, the scaling structure of avalanche behavior is lost, and the
growth process produces either extremely large or very small
avalanches. This feature makes it prohibitively difficult to take this
regime under closer scrutiny with numerical simulations. The time to
reach $m_{cr}$ can be estimated in two ways depending on whether one
uses macroscopic time or 'SOC' time (counting the number of
avalanches). With a uniformly, but very slowly increasing force $F$
the critical slope is reached at the same time regardless of the
system size $L$. In contrast, since the average size of an avalanche
is independent of $L$ (recall that $\tau_s > 2$), it follows that the
number of avalanches $n_{cr}$ to reach $m_{cr}$ will depend on $L$ as
$n_{cr} \sim L^2$.

The previous results illustrate the nature of SOC in the QKPZ
universality class. The combination of slow drive and boundary pinning
chooses a certain symmetry for the average interface profile. This can
be further elaborated by considering the continuum equation, and
writing the interface field on one of the sides of the triangle as
$H(x,t) = m(t)x + \delta H(x,t)$. Inserting this into the QKPZ
equation we obtain for $\delta H$
\begin{eqnarray}
{\partial \delta H({\mathbf x}, t) \over \partial t} = \nu \nabla^2
\delta H + {\lambda} m(t) \nabla \delta H + {\lambda \over 2} (\nabla
\delta H)^2
\nonumber \\
+ \eta({\mathbf x}, H),
\label{mKPZ}
\end{eqnarray}
where the mean-field solution for $H$ has been subtracted from the
equation and the result is valid for intermediate times $\delta t \ll
t$.  We discover that the externally imposed slope (by the combination
of $F$ and $H=0$ at the boundaries) {\em changes the effective
interface equation} by creating a term linear in $\nabla \delta H$,
that will dominate over the KPZ-nonlinearity. The situation is thus
exactly analogous to depinning transitions in the presence of an
anisotropy, discussed by Kardar, Dhar, and Tang \cite{kdt}. This
explains qualitatively why the avalanche behavior at criticality, at
the SOC depinning transition, is different from that of the bulk
depinning transition for the original equation. Regardless of whether
the SOC ensemble differs from the translationally invariant one for
ordinary SOC sandpiles, it is thus clear that in the case of SOC
obtained with a similar recipe in QKPZ-systems the two ensembles do
not share the same exponents. Should it be possible to write an
activity-based description for the QKPZ universality class, this
implies that the field theory should have a different structure in the
two cases, in contrast to the connection between the QEW and absorbing
state phase transitions with a conserved field \cite{munoz}.

Above, we have concentrated on a cellular automaton for the QKPZ
equation since as discussed it is related to a microscopic 'activity'
picture, or a ``QKPZ sandpile''.  The slow-drive SOC-depinning limit
can be taken also in the continuum version. Our investigations
indicate that the same phenomenology of symmetry breaking persists in
that case as well, though the actual exponents in the second phase
before the critical slope is reached differ as $\Delta h$ can now have
arbitrary values and the avalanche properties change.  The analysis of
the continuum model is hampered by well-known numerical instabilities
that develop once the smoothing effect of the Laplacian is reduced -
which is inevitable here due to the increasing average slopes.  We
have also studied the two-dimensional case. While there, with periodic
boundary conditions in one direction, and SOC ones in the other, are
interesting details concerning the geometry of avalanche spreading,
the basic physics is the same as in 1D.

In conclusion, in this paper we have analyzed the behavior of
interfaces in the QKPZ universality class in the presence of
conditions that produce self-organized criticality. Spatial and
temporal invariance are broken by the slow drive and the dissipative
pinning boundary conditions, giving rise to a macroscopic interface
shape.  The shape, in combination with the nonlinearity, results in
the creation of a term linear in $\delta h$ in the effective equation
that governs the physics of avalanches. In contrast to usual SOC
sandpiles we find that there are two avalanche regimes depending on
the microscopic dynamics, where the second one fails to show critical
scaling seen in the first. The ensemble in which the criticality
appears is fundamentally different from the usual depinning one and is
related to the anisotropic QKPZ class \cite{kdt}.

Partial support of OTKA T029985 is acknowledged.

\end{multicols}

\end{document}